\documentclass[10pt,journal,compsoc]{IEEEtran}
\pdfoutput=1
\ifCLASSOPTIONcompsoc
  \usepackage[nocompress]{cite}
\else
  \usepackage{cite}
\fi

\ifCLASSINFOpdf
\else
\fi

\usepackage[pdftex]{hyperref}

\usepackage[ruled]{algorithm2e}
\usepackage{my}
\graphicspath{{./images/}}

\usepackage{color}
\definecolor{MainColor}{RGB}{128, 0, 128}

\hypersetup{
    unicode=true,          
    pdftoolbar=true,        
    pdfmenubar=true,        
    pdffitwindow=false,     
    pdfstartview={FitH},    
    pdftitle={Voting for Distortion Points in Geometric Processing},    
    pdfnewwindow=true,      
    colorlinks=true,       
    linkcolor=MainColor,          
    citecolor=MainColor,        
    filecolor=magenta,      
    urlcolor=MainColor,           
    breaklinks=true
}

\begin{document}

\title{Voting for Distortion Points in\\Geometric Processing}

\author{Shuangming~Chai, Xiao-Ming~Fu, and Ligang~Liu
\IEEEcompsocitemizethanks
{
\IEEEcompsocthanksitem The authors are with the School of Mathematical Sciences,
University of Science and Technology of China, Hefei, Anhui 230026, China.\protect\\
E-mail: kfckfckf@mail.ustc.edu.cn, \{fuxm, lgliu\}@ustc.edu.cn.
}
\thanks{Corresponding author: Xiao-Ming~Fu.}
}


\IEEEtitleabstractindextext{%
\begin{abstract}
Low isometric distortion is often required for mesh parameterizations.
A configuration of some vertices, where the distortion is concentrated, provides a way to mitigate isometric distortion, but determining the number and placement of these vertices is non-trivial.
%
We call these vertices \emph{distortion points}.
We present a novel and automatic method to detect distortion points using a voting strategy.
Our method integrates two components: candidate generation and candidate voting.
Given a closed triangular mesh, we generate candidate distortion points by executing a three-step procedure repeatedly: (1) randomly cut an input to a disk topology; (2) compute a low conformal distortion parameterization; and (3) detect the distortion points.
Finally, we count the candidate points and generate the final distortion points by voting.
We demonstrate that our algorithm succeeds when employed on various closed meshes with a genus of zero or higher.
The distortion points generated by our method are utilized in three applications, including planar parameterization, semi-automatic landmark correspondence, and isotropic remeshing.
Compared to other state-of-the-art methods, our method demonstrates stronger practical robustness in distortion point detection.
\end{abstract}

}

\maketitle
\IEEEdisplaynontitleabstractindextext
\IEEEpeerreviewmaketitle

\section{Introduction} \label{sec:intro}
\begin{figure*}
  \centering
  \begin{overpic}[width=0.95\linewidth]{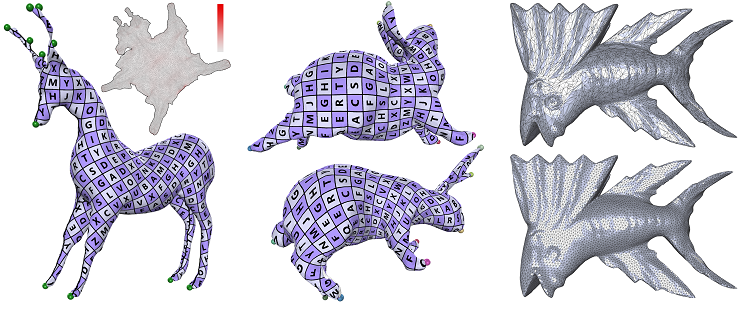}
    \put( 8,0.5){\small \textbf{Planar Parameterization} }
    \put(40,0.5){\small \textbf{Surface Correspondence} }
    \put(75,0.5){\small \textbf{Isotropic Remeshing} }
    \put(30.5,36){\scriptsize 1.0}
    \put(30.5,42.5){\scriptsize 5.0}
  \end{overpic}
  \caption{
    We employ the detected distortion points (colorized spheres) in three applications (See more details in Section~\ref{sec:app}).
    Note that in this paper, the triangles are colored in red with different saturations according to the isometric distortion ranging from 1.0 to 5.0, as shown in the left figure.
  }
  \label{fig:teaser}
\end{figure*}

\IEEEPARstart{L}{ow} isometric distortion parameterization plays a fundamental role in many computer graphics and geometric processing tasks, such as atlas generation~\cite{poranne2017autocuts,LiuAtlas2018}, remeshing~\cite{botsch2004remeshing,bommes2009mixed}, and bijective surface correspondence~\cite{Aigerman2014}.
Fig.~\ref{fig:teaser} shows three applications.
Before being parameterized to the plane, a closed mesh needs to be cut to a disk topology.
The distortion of a parameterization is highly related to not only the parameterization method but also the position of the cut.
Since a surface is generally not developable, the isometric parameterization distortion is usually high, unless the cut passes through some vertices, where the distortion is usually concentrated.
We call these vertices \emph{distortion points}.

A \emph{good} distortion point detection algorithm usually satisfies the following properties:
(1) the algorithm is automatic and does not require manual intervention;
(2) the parameterizations exhibit low isometric distortion if the cut path passes through the detected distortion points;
(3) the number of distortion points is small;
(4) and the algorithm is efficient, otherwise manual selection is preferable.
On one hand, more distortion points tend to reduce isometric distortion, but too many distortion points are very likely to have a negative impact on subsequent applications.
For example, more distortion points usually lead to longer cut, which may worsen the rendering performance~\cite{Hakura:1997} and increase the texturing artifacts~\cite{poranne2017autocuts}.
On the other hand, fewer distortion points usually produce higher isometric distortion.
Therefore, there is a trade-off between properties (2) and (3).

\begin{figure}[!t]
  \centering
  \begin{overpic}[width=\linewidth]{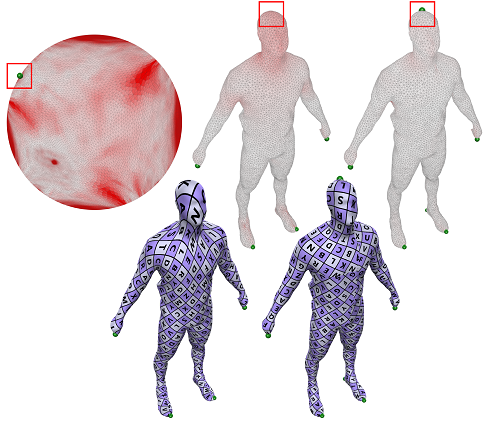}
    \put(17,40){\small \textbf{(a)}}
    \put(37,0.5){\small \textbf{(b)}}
    \put(53,35){\small \textbf{(c)}}
    \put(67,0.5){\small \textbf{(d)}}
    \put(85,35){\small \textbf{(e)}}
    \put( 0,10){\scriptsize \parbox{0.5\linewidth}{
    \begin{tabular}{l}
      \textbf{Geometry Images~\cite{Gu:2002}}\\
      \midrule
      $\delta_\text{avg}=1.29$\\
      $\delta_\text{max}=3.24$\\
      $\delta_\text{std}=0.32$
    \end{tabular}
    }}
    \put(78,10){\scriptsize \parbox{0.5\linewidth}{
    \begin{tabular}{l}
      \textbf{Ours}\\
      \midrule
      $\delta_\text{avg}=1.11$\\
      $\delta_\text{max}=4.71$\\
      $\delta_\text{std}=0.12$
    \end{tabular}
    }}
  \end{overpic}
  \caption{
    An example of a defect of distortion-based methods~\cite{Gu:2002,Springborn2008}.
    If the maximum distortion is on the boundary vertex in one iteration (a), then there is no new cut added, and the iterative process stops.
    This leads to missing features, i.e. the head of the man (b \& c), and high isometric distortion (as shown in the red box in (c)).
    Our method finds the distortion points correctly (d \& e), and the distortion is much lower.
  }
  \label{fig:distortion}
\end{figure}

\begin{figure}[!t]
  \centering
  \begin{overpic}[width=\linewidth]{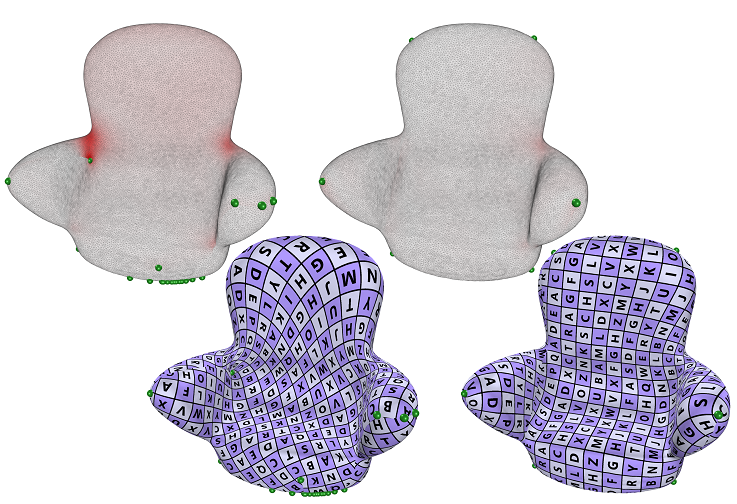}
    \put( 0,20){\scriptsize \parbox{0.5\linewidth}{
    \begin{tabular}{l}
      \textbf{Seamster~\cite{sheffer2002seamster}}\\
      \midrule
      $\delta_\text{avg}=1.35$\\
      $\delta_\text{max}=12.6$\\
      $\delta_\text{std}=0.37$
    \end{tabular}
    }}
    \put(78,50){\scriptsize \parbox{0.5\linewidth}{
    \begin{tabular}{l}
      \textbf{Ours}\\
      \midrule
      $\delta_\text{avg}=1.13$\\
      $\delta_\text{max}=3.75$\\
      $\delta_\text{std}=0.10$
    \end{tabular}
    }}
  \end{overpic}
  \caption{
    An example of a drawback of the curvature-based method~\cite{sheffer2002seamster}.
    Since the curvature is related to the mesh triangulation, if the mesh has high resolution, then the curvature-based method finds redundant high-curvature points and misses important features, such as the backrest of the armchair (left).
    Our method correctly generates the result with lower distortion (right).
  }
  \label{fig:curvature}
\end{figure}

From manual picking to automatic generation, there have been many attempts to detect distortion points~\cite{Springborn2008,Gu:2002,ChaiCut2018}.
Since isometric distortion is the key to our problem, a possible strategy for detecting distortion points is to directly use isometric distortion as a core measure.
For instance, some methods iteratively parameterize the mesh to the plane and add the point of greatest isometric distortion into the distortion point set~\cite{Gu:2002,Springborn2008}.
However, the point of greatest distortion is often located at a boundary, thereby leading to early termination and causing high isometric distortion (see Fig.~\ref{fig:distortion}).
Instead of using planar parameterizations, Chai~\etal~\cite{ChaiCut2018} cluster high isometric distortion points using an as-conformal-as-possible (ACAP) spherical parameterization of the input mesh~\cite{Hu-2017-AHSP}.
Although the clustering technique used in this method is very effective for detecting distortion points, it remains a challenge to efficiently compute bijective spherical parameterizations with theoretical guarantees.

In this paper, we propose a novel method to detect distortion points.
Due to recent developments in planar parameterization techniques~\cite{Tutte1963,Shen2019}, a flip-free planar parameterization is more computationally efficient and robust than a spherical parameterization.
Thus, we hope to use planar parameterizations instead of spherical parameterizations to find distortion points efficiently and robustly.
However, it is difficult due to the following two reasons.
First, planar parameterizations require that the input closed mesh is cut to a disk topology.
However, since different cuts result in different parameterizations, the cut location has a significant effect on the detection result.
Hence how to prepare a cut and handle large distortion near boundaries is very challenging.
Second, robustly identifying distortion points based on planar parameterizations is non-trivial.

To eliminate the effects of different cuts, we present a simple and effective voting strategy to detect distortion points.
To generate candidate distortion points to vote on, we repeatedly perform the following three-step procedure: first cut the input mesh randomly, then compute a low conformal distortion planar parameterization, and finally detect a set of vertices as candidates.
To consistently seek candidates with strong practical robustness, we customize an effective clustering strategy, shown in~\cite{ChaiCut2018}, which uses a parameter to control the balance between the number of distortion points and the final parameterization distortion.
Note that the generated candidates may include some vertices near boundaries, but they are usually detected only once since the random cut paths have almost no intersections.
Therefore, in the voting phase, we only recognize vertices that appear more than once as the final candidates.

Our approach is simple and performs better than current state-of-the-art methods.
We conduct various experiments and comparisons that demonstrate the feasibility and effectiveness of our method.
Furthermore, we successfully employ the detected distortion points in three applications, including low-distortion parameterizations, semi-automatic corresponding landmark detection for bijective surface mappings, and isotropic remeshing (Fig.~\ref{fig:teaser}).

\section{Related Work} \label{sec:related}

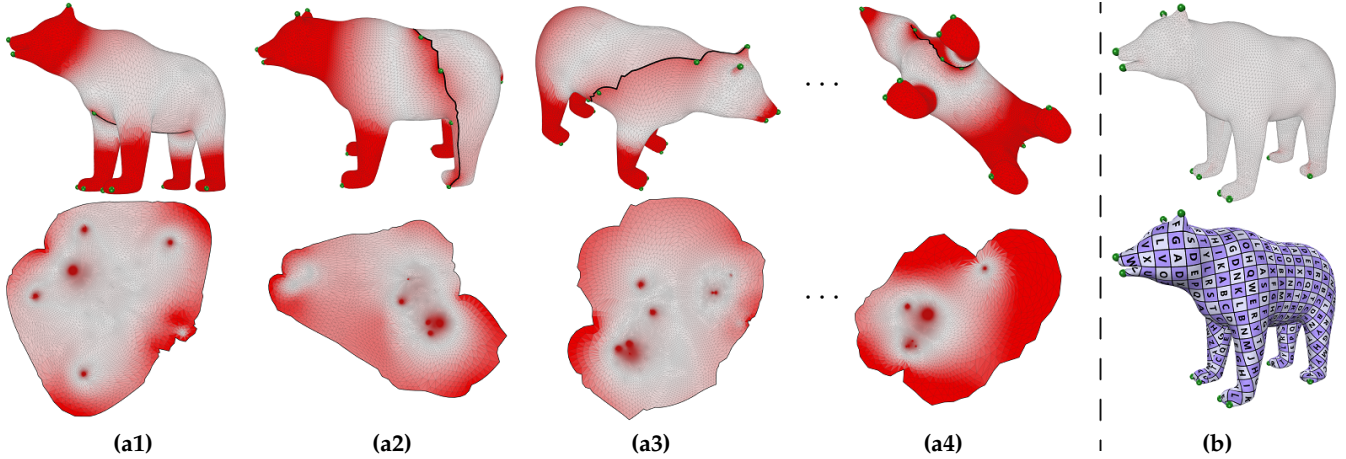
\begin{figure*}[t]
  \centering
  \begin{overpic}[width=0.97\linewidth]{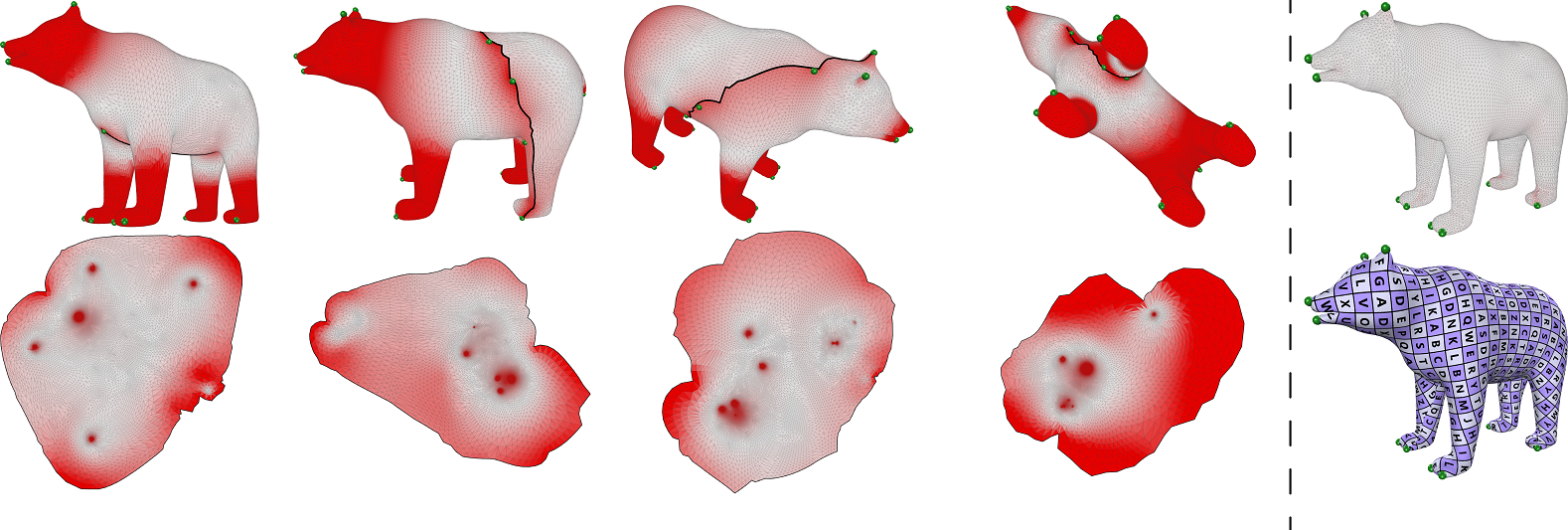}
    \put(8,0){\small\textbf{(a1)} }
    \put(27.5,0){\small\textbf{(a2)} }
    \put(47,0){\small\textbf{(a3)} }
    \put(69,0){\small\textbf{(a4)} }
    \put(60,11){\large $\cdots$}
    \put(60,27){\large $\cdots$}
    \put(90,0){\small\textbf{(b)} }
  \end{overpic}
  \caption{
    The overview of our algorithm.
    We cut the mesh randomly, parameterize it to a disk as conformally as possible (a1--a4 bottom), and cluster the candidate feature points (green points in a1--a4 along the top).
    Judging from the candidate points, we select the final feature points according to a voting strategy (b).
    The parameterization results are generated by~\cite{LiuPP-2018}.
  }
  \label{fig:pipeline}
\end{figure*}

\emph{Distortion Point Detection.}
Since vertices with high curvatures often induce high isometric distortion, some algorithms try to detect distortion points by using the curvature information as a guide~\cite{sheffer2002,sheffer2002seamster}.
However, as curvature-based methods do not consider distortion directly, some distortion points may be ignored (see Fig.~\ref{fig:curvature}).
Distortion measures are used to develop an iterative algorithm that alternately flattens the mesh and puts the maximum distortion vertex into the distortion point set~\cite{Gu:2002}.
However, as mentioned in the previous section and Fig.~\ref{fig:distortion}, it is difficult to avoid early termination due to the maximum distortion vertex's location on the boundary.
To overcome this issue, spherical parameterizations are used~\cite{ChaiCut2018}; however, efficiently computing bijective spherical parameterizations for complex objects is a non-trivial task (see the comparison in Fig.~\ref{fig:sphere}).

\emph{Conformal Cone Singularities.}
In the context of conformal flattening, finding cone singularities is an effective way to mitigate isometric distortion.
Kharevych~\etal~\cite{kharevych2006} manually place cone singularities and adjust cone angles to reduce distortion.
Similar to~\cite{Gu:2002}, an iterative algorithm is used in~\cite{Springborn2008} to compute cone configurations, including the number, placement, and size.
Curvatures are also used to find the cone singularities~\cite{ben2008conformal,Myles2012,vintescu2017}; however, curvatures do not always provide useful information about how cones should be arranged, as observed in~\cite{Soliman:2018}.
Thus, starting from a different perspective, Soliman~\etal~\cite{Soliman:2018} study the problem of finding optimal cones as an approximation problem.
However, their optimal cone configurations are affected by a parameter that has no intuitive geometric interpretation.
When using default values, some important vertices are missed (see Fig. 15 in~\cite{Soliman:2018}).
In our method, the trade-off between the number of distortion points and the level of isometric distortion is controlled by a parameter that represents the range of influence of distortion points.

\emph{Singularities of Fields.}
Some recent quadrilateral remeshing methods~\cite{bommes2009mixed,Bommes:quadsurvey2013} cut the input mesh to a disk topology by adding the singularities of the designed fields into the cut paths.
The field design procedure~\cite{vaxman2016} and parameterization computation step are separate, so the connections between the field singularities and parameterization distortion are not close and indirect.
As shown in Fig. 17 in~\cite{Soliman:2018}, the highly regular fields generated by~\cite{Knoppel:2013} do not immediately result in a low isometric distortion parameterization.
Thus, the isometric distortion of parameterizations may still be large, even if the singularities are from highly smooth fields.

\emph{Cut Construction and Atlas Generation.}
Instead of defining cuts by connecting vertices, some methods directly construct cut paths to generate low isometric distortion parameterizations.
Mesh segmentation approaches~\cite{Levy2002,Sander:2002,Sander2003,Zhou2004,julius2005d,zhang2005} partition an input mesh into multiple charts, and each chart is parameterized with very low isometric distortion.
Generating a quad layout~\cite{campen2017partitioning} is an another way to segment an input mesh into multiple charts, such as the domain simplicity~\cite{Tarini:2011}, aligned global parameterizations~\cite{Campen2014QLE}, perfect matching~\cite{Razaflndrazaka:2015,Razafindrazaka201763}, reliable quad meshing~\cite{Bommes:2013_IGM}, and field tracing~\cite{Campen:2012,Pietroni2016}.
Techniques~\cite{Sorkine2002,poranne2017autocuts} that simultaneously cut and flatten input meshes have also been developed.
Li~\etal~\cite{Li:2018:OptCuts} propose an algorithm to jointly optimize the parameterization and the cut-preserving bijection of the mapping.
In contrast to these methods, our goal is to automatically detect distortion points, and use them in applications beyond parameterization generation.
Liu~\etal~\cite{Liu2019} generate cuts for peeling art.

\section{Voting for distortion points} \label{sec:method}

\subsection{Overview}\label{sec:Overview}
We study distortion point detection on 3D surface models.
Given a triangular mesh $\mM$ containing $N_v$ vertices $\mV=\{\mv_i,i=1,\ldots,N_v\}$ and $N_t$ triangles $\mT=\{\mt_i,i=1,\ldots,N_t\}$, our goal is to find a set of distortion points $\mV^d$.
To achieve this goal, we propose a voting-based algorithm.
In Section~\ref{sec:candidate}, we present a three-step method to generate candidate distortion points.
After performing the three-step procedure many times, we develop a voting strategy to determine the resulting distortion points, as described in Section~\ref{sec:voting}.
Some implementation details and discussions are provided in Section~\ref{sec:details}.
Fig.~\ref{fig:pipeline} illustrates the pipeline of our method using a bear model.

\subsection{Candidate Distortion Points}\label{sec:candidate}
Instead of spherical parameterizations~\cite{ChaiCut2018}, we use distortion measures on planar parameterizations to detect distortion points.
Thus, we first cut $\mM$ as a disk topology mesh $\mM^c$ (Section~\ref{sec:cut}), then parameterize $\mM^c$ onto the plane (Section~\ref{sec:para}), and finally find the distortion points (Section~\ref{sec:cluster}).

\begin{figure}
\centering
\begin{overpic}[width=\linewidth]{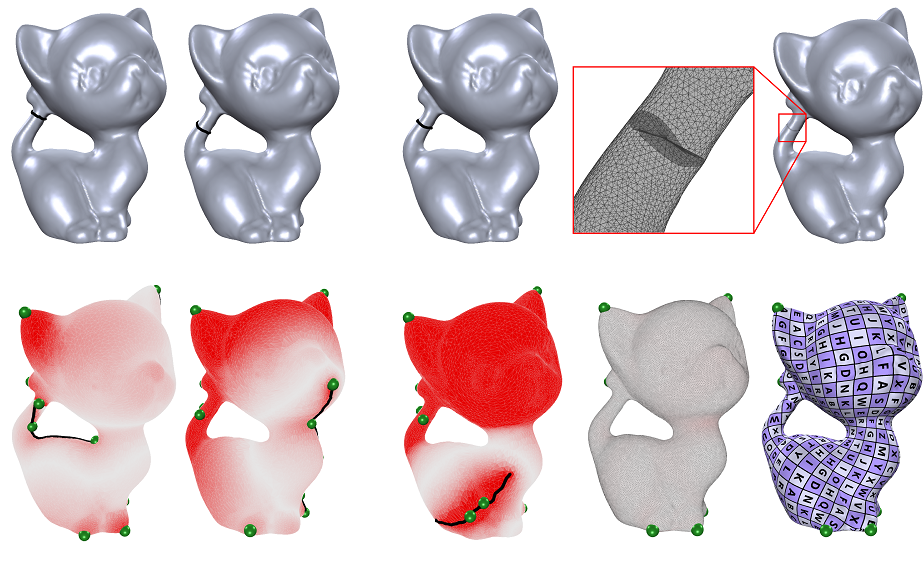}
\put(6,32.5){\small\textbf{(a1)}}
\put(26,32.5){\small\textbf{(a2)}}
\put(49,32.5){\small\textbf{(a3)}}
\put(89,32.5){\small\textbf{(b)}}
\put(6,0){\small\textbf{(c1)}}
\put(26,0){\small\textbf{(c2)}}
\put(49,0){\small\textbf{(c3)}}
\put(68,0){\small\textbf{(d1)}}
\put(87,0){\small\textbf{(d2)}}
\put(38,48){$\cdots$}
\put(38,16){$\cdots$}
\end{overpic}
\caption{
Steps of our algorithm for high-genus meshes.
We first generate handles and randomly perturb them (a1--a3), then cut along the handles (b) and fill the holes (see the zoom-in).
By applying our algorithm for genus-zero meshes, we obtain some candidate distortion points (c1--c3).
After the voting process, the final distortion points are detected (d1--d2).
}
\label{fig:highpipeline}
\end{figure}
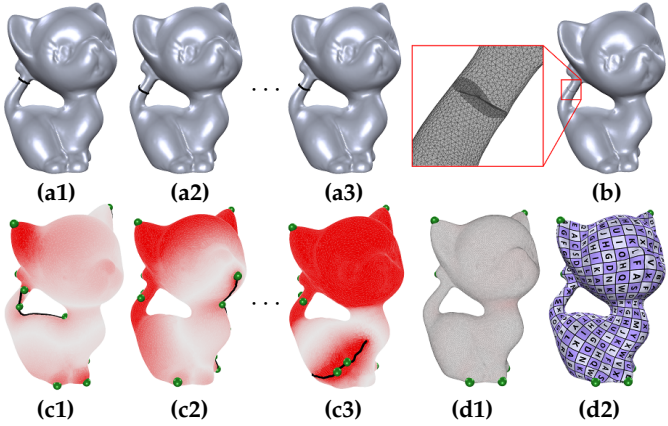

\subsubsection{Constructing cuts}\label{sec:cut}
\emph{Genus-Zero Surfaces.}
We generate cuts for genus-zero surfaces as follows:
\begin{enumerate}
  \item Randomly select a vertex $\mv_i$;
  \item Find the farthest vertex $\mv_j$ from $\mv_i$, i.e., $\mv_j$ satisfies $\|\mv_i-\mv_j\|^2 \geq \|\mv_i-\mv_k\|^2$, for all $k \in \{1,\ldots,N_v\}$, where $\|\cdot\|$ refers to the Euclidean norm;
  \item Compute the shortest path between $\mv_i$ and $\mv_j$ along the mesh edges as a cut.
\end{enumerate}

\emph{High-Genus Surfaces.}
For high-genus surfaces, we first convert them into genus-zero surfaces and then find the cut using the above procedure for genus-zero surfaces.
The conversion from high genus to zero genus is as follows:
\begin{enumerate}
  \item Compute the handles of the high-genus surface using~\cite{Dey:2013};
  \item Randomly perturb the generated handles by adding a random offset in terms of the geodesic distance;
  \item Cut the high-genus surface along these perturbed handles to create some holes;
  \item Fill in the holes to generate a genus-zero surface.
\end{enumerate}
Fig.~\ref{fig:highpipeline} uses an example to show the distortion detection process for high-genus models.
In general, vertices that are located on handles are usually identified as distortion points using our detection algorithm.
To distinguish them from true distortion points, we randomly perturb the generated handles of~\cite{Dey:2013} so that the vertices on the handles become different.
After the voting process, the vertices caused by the handles are filtered out since they only appear once.

\subsubsection{Parameterization of $\mM^c$}\label{sec:para}
Inspired by methods~\cite{Gu:2002,Springborn2008} that iteratively compute a conformal parameterization and place a distortion point at the vertex with the highest isometric distortion, our method also generates a parameterization that is as conformal as possible with no flips for detecting distortion points.
Numerous techniques for ACAP parameterizations have been developed (cf. the surveys in~\cite{Floater2005,Sheffer2006,Hormann:2007}).
To achieve a flip-free and ACAP conformal parameterization, we initialize the parameterization using Tutte's embedding method and optimize the AMIPS energy~\cite{Fu2015} to reduce the conformal distortion.

\emph{Formulation.}
A parameterization $\Phi$ of $\mM^c$ is constituted by piecewise constant affine maps $\phi_{\mt}$ defined on triangles $\mt \in \mM^c$.
After defining a coordinate system on each triangle $\mt$, each affine map $\phi_{\mt}$ has a linear form $\phi_{\mt}(\mx) = J_{\mt} \mx + \mb_{\mt}$, where $J_{\mt}$ refers to the Jacobian of $\Phi$.
The MIPS energy~\cite{Hormann2000} is defined as follows:
\begin{equation}\label{equ:MIPS}
  E_\text{MIPS}(J_{\mt}) = \frac{1}{2} \left(\frac{\sigma_1}{\sigma_2} + \frac{\sigma_2}{\sigma_1}\right) = \frac{1}{2} \frac{\| J_{\mt} \|_F^2}{\det J_{\mt}},
\end{equation}
where $\sigma_1$ and $\sigma_2$ are singular values of $J_{\mt}$, and $\|\cdot\|_F$ is the Frobenius norm.
$E_\text{MIPS}(J_{\mt})$ reaches its minimum when $J_{\mt}$ represents a similarity transformation.
The AMIPS energy~\cite{Fu2015} is defined as follows:
\begin{equation}\label{equ:AMIPS}
  \begin{aligned}
    E_\text{AMIPS}(J_{\mt}) &= \exp (E_\text{MIPS}(J_{\mt})), \\
    E_\text{AMIPS}  &= \sum_{\mt \in \mM^c} E_\text{AMIPS}(J_{\mt}).
  \end{aligned}
\end{equation}
Optimizing $E_\text{AMIPS}$ penalizes the maximum distortion and generates evenly and smoothly distributed distortion.

\begin{figure}[t]
  \centering
  \begin{overpic}[width=\linewidth]{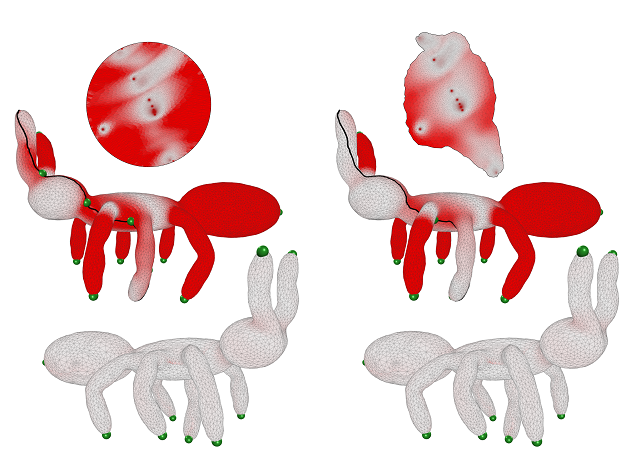}
    \put(12,68){\small \textbf{Fixed Boundary}}
    \put(62,68){\small \textbf{Free Boundary}}
  \end{overpic}
  \caption{
    Comparison of different types of parameterizations.
    It takes 1.28 and 0.84 seconds to compute a parameterization while fixing the boundary and relaxing the boundary, respectively.
    Compared to free boundary parameterizations, fixed boundary parameterizations generate similar distortion distributions (top and middle) and find the same final feature points (bottom), but fixed boundary parameterizations take more time to optimize distortions.
  }
  \label{fig:para_type}
\end{figure}

\emph{Optimization.}
Unfortunately, conformal parameterizations are not unique.
For example, a new parameterization $\widehat{\Phi} \coloneqq f \circ \Phi$, where $f$ is an arbitrary M\"{o}bius transformation, contains the same conformal distortion as $\Phi$.
However, the area distortion of $\widehat{\Phi}$ and $\Phi$ are likely to be very different.
Thus, if the generated ACAP parameterizations are not appropriate, the vertices with high isometric distortion may be located incorrectly.
These conditions would affect our technique's ability to robustly detect distortion points.

In our method, $\Phi$ is initialized by Tutte's embedding method~\cite{Tutte1963}.
As observed in our experiments, if we try to minimize the MIPS and AMIPS energy while fixing the boundary vertices, the vertices with high isometric distortion often become our desired distortion points.
We also test free boundary cases and observe that the distortion distribution is similar except for the distortion near boundaries. (see the comparison in Fig.~\ref{fig:para_type}).
However, the running time of free boundary parameterization is faster.
Thus, the boundary vertices are always free during the conformal distortion optimization in our experiments, which use the KP-Newton method~\cite{golla2018} to optimize $E_\text{AMIPS}$.

\subsubsection{Detecting Distortion Points}\label{sec:cluster}
\emph{Isometric Distortion.}
After achieving an ACAP parameterization $\Phi$, we select the vertices with high isometric distortion as our distortion points.
The isometric distortion~\cite{Fu2015} is defined as follows:
\begin{equation}\label{equ:area}
  \begin{aligned}
    E_\text{area}(J_{\mt}) & = \frac{1}{2}(\det J_{\mt} + \det J^{-1}_{\mt}), \\
    E_\text{iso}(J_{\mt})  & = \frac{1}{2}(E_\text{area}(J_{\mt}) + E_\text{MIPS}(J_{\mt})).
  \end{aligned}
\end{equation}
If $J_{\mt}$ is a rotation matrix, $E_\text{iso}(J_{\mt})$ reaches its minimum.

\IncMargin{0.5em}
\begin{algorithm}[t]
\caption{Distortion Triangle Detection}
\label{alg:dtd}
\SetCommentSty{textsf}
\SetKwInOut{AlgoInput}{Input}
\SetKwInOut{AlgoOutput}{Output}

\SetKwFunction{GCT}{GroupTri2Region}
\SetKwFunction{TNR}{TriNumber}
\SetKwFunction{PF}{PopFront}
\SetKwFunction{PB}{PushBack}

\AlgoInput{ An ACAP parameterization $\Phi$ }
\AlgoOutput{ A set of distortion triangles $\mT^d$ }
Compute $E_\text{iso}(J_{\mt}),\forall \mt \in \mT$\;
Initialize a queue $Q\gets\{\mT\}$, and $\mT^d\gets\emptyset$\;
\While {$Q \neq \emptyset$}
{
    $\widehat{\mT} \leftarrow \PF(Q)$\tcp*{Pop the front of $Q$}
    $\widehat{\mt} \leftarrow \argmax_{\mt \in \widehat{\mT}} \{E_\text{iso}(J_{\mt}) | \mt \in \widehat{\mT}\} $\;
    \If {$\widehat{\mt} \not \in \mT^d$}
    {
        $\mT^d\gets\mT^d\cap\{\widehat{\mt}\}$\;
    }
    $E_\text{iso}^\text{median} \leftarrow \median\{E_\text{iso}(J_{\mt}) | \mt \in \widehat{\mT}\} $\tcp*{Find the median of isometric distortions on the triangles of $\widehat{\mT}$}
    $\widetilde{\mT}\gets\{\mt\in\widehat{\mT}|E_\text{iso}(J_{\mt}) \geq E_\text{iso}^\text{median} \}$\;
    $R \leftarrow \GCT(\widetilde{\mT})$\tcp*{Group the connected triangles into several new isolated regions}
    \For{$R_i\in R$}
    {
        \If(\tcp*[f]{$N$ is a threshold}){$\TNR(R_i) \geq N$}
        {
            $\PB(Q,R_i)$\tcp*{Push $R_i$ at the back of $Q$}
        }
    }
}
\end{algorithm}
\DecMargin{0.5em}

\emph{Clustering Method.}
Since the clustering method proposed in~\cite{ChaiCut2018} is practically robust for detecting distortion points $\mV^d$, we first review it and then modify it for our ACAP planar parameterizations.

In~\cite{ChaiCut2018}, the distortion triangles $\mT^d$ each have local maximum distortion over a region.
These triangles are first detected, and then one vertex from each distortion triangle is added to $\mV^d$.
The pseudocode of the detection algorithm for $\mT^d$ is listed in Algorithm~\ref{alg:dtd}.
The threshold $N$ indicates the distortion point scale (see a more detailed discussion in Section~\ref{sec:details}).

\begin{figure}[t]
  \centering
  \begin{overpic}[width=\linewidth]{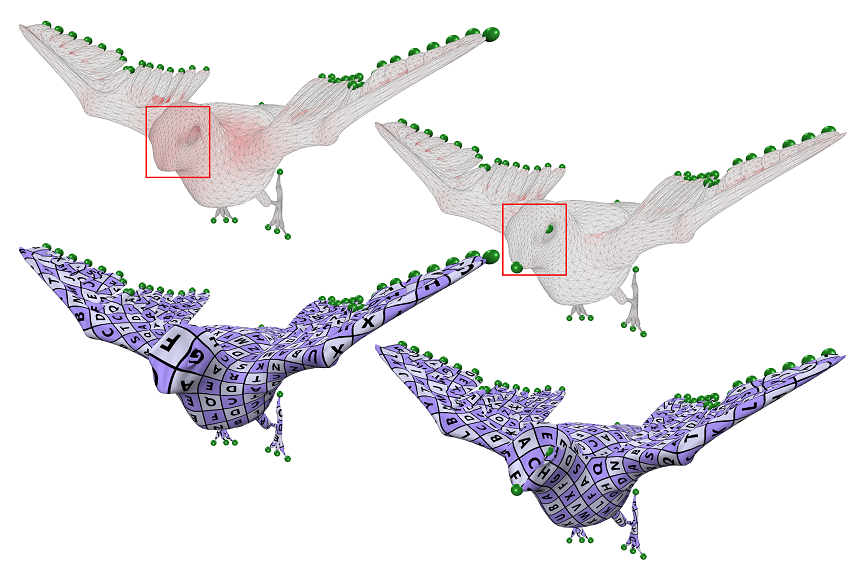}
    \put( 0, 7){\scriptsize \parbox{0.5\linewidth}{
    \begin{tabular}{l}
      $\delta_\text{avg}=1.29$\\
      $\delta_\text{max}=4.55$\\
      $\delta_\text{std}=0.32$\\
      \midrule
      \textbf{No first filtering}
    \end{tabular}
    }}
    \put(70,57){\scriptsize \parbox{0.5\linewidth}{
    \begin{tabular}{l}
      \textbf{With first filtering}\\
      \midrule
      $\delta_\text{avg}=1.17$\\
      $\delta_\text{max}=4.79$\\
      $\delta_\text{std}=0.22$
    \end{tabular}
    }}
  \end{overpic}
  \caption{
    Comparison of a modification in the clustering process.
    If we filter half of the triangles directly in the first iteration, some feature points in the red box are missing (left).
    If we first filter the triangles whose area distortions are smaller than a threshold $E^\text{th}_\text{iso}$, then there will be more feature points detected (right).
  }
  \label{fig:frist_K}
\end{figure}

In the first iteration, since $\widehat{\mT}$ contains all of the triangles, the algorithm adds the triangle with maximum distortion into $\mT^d$ and filters out half of the triangles.
For some models with ACAP planar parameterizations, some distortion triangles may be filtered out by the median (see Fig.~\ref{fig:frist_K} left).
To avoid this issue, we modify the clustering method by choosing a different strategy for the first iteration.
Since our goal is to detect the triangles at the local maxima, the isometric distortion in our target triangles is at least greater than the small value $E_\text{iso}^\text{th}$.
In the first iteration, we filter out the triangles with isometric distortions that are smaller than $E_\text{iso}^\text{th}$.
Then, we group the remaining triangles into new isolated regions to initialize the queue $Q$.
Finally, we run the ``while'' loop of Algorithm~\ref{alg:dtd} to compute $\mT^d$ using the median filter.
With this simple modification, our method catches obvious distortion triangles, resulting in parameterizations with lower isometric distortion (see Fig.~\ref{fig:frist_K} right).
In our experiments, we set $E_\text{iso}^\text{th} = 2$ in all of the models.

\subsection{Voting}\label{sec:voting}
We present a simple voting strategy to generate final distortion points.
This strategy is based on the fact that no matter how cuts change, some vertices are almost always detected as distortion points.
The voting strategy contains two steps:
(1) perform the candidate distortion point generation procedure many times (ten in our experiments);
and (2) vote for the candidates that appear multiple times (more than twice in our experiments) to serve as the resulting distortion points.
Fig.~\ref{fig:pipeline}b shows the voting results from the candidates shown in Figs.~\ref{fig:pipeline}a1, \ref{fig:pipeline}a2, \ref{fig:pipeline}a3, and \ref{fig:pipeline}a4).

\begin{figure}[t]
  \centering
  \begin{overpic}[width=\linewidth]{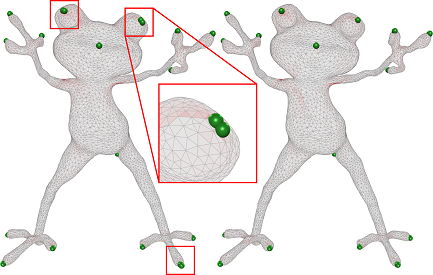}
    \put(9,0){\small \textbf{No post-filtering} }
    \put(59,0){\small \textbf{With post-filtering} }
  \end{overpic}
  \caption{
    Results of our algorithm with and without the use of post-filtering.
    The post-filtering is used to detect redundant distortion points, which are very close to each other and considered part of the same distortion point, marked in the red box (left).
    We remove all the other points in each small neighborhood, while leaving one point that receives the most votes as the final distortion point (right).
  }
  \label{fig:post}
\end{figure}

\emph{Post-Filtering.}
Due to numerical issues, some distortion points are very close to each other.
Grouping these close distortion points into one point does not affect the resulting isometric distortion (see the comparison in Fig.~\ref{fig:post}).
If one distortion point $\mv^d_i$ is within the $n$-ring neighbors of another distortion point $\mv^d_j$, we keep the one that has more votes and discard the other one.
In the experiments, we observe that the very close points always lie in one-ring or two-ring neighborhoods, so we set $n$ as 5 to filter out close points effectively.

\begin{figure}[t]
  \centering
  \begin{overpic}[width=\linewidth]{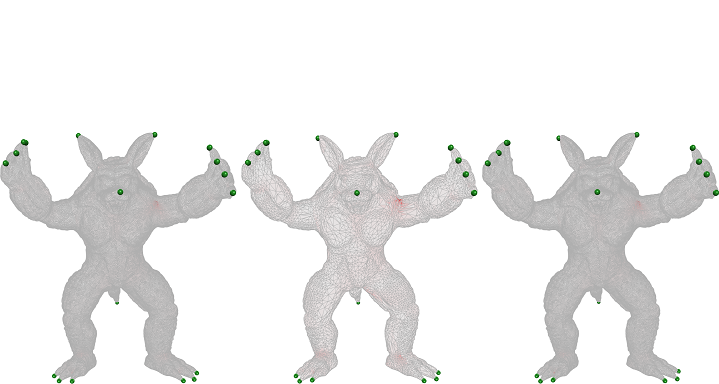}
    \put(3,43){\scriptsize \parbox{0.5\linewidth}{
    \begin{tabular}{l}
      \textbf{No Simplification}\\
      \midrule
      $\delta_\text{avg}=1.19$\\
      $\delta_\text{max}=4.84$\\
      $\delta_\text{std}=0.12$\\
    \end{tabular}
    }}
    \put(46,43){\scriptsize \parbox{0.5\linewidth}{
    \begin{tabular}{l|l}
      \multicolumn{2}{c}{\textbf{With Simplification}}\\
      \midrule
      $\delta_\text{avg}=1.20$ & $\delta_\text{avg}=1.19$\\
      $\delta_\text{max}=7.75$ & $\delta_\text{max}=4.84$\\
      $\delta_\text{std}=0.15$ & $\delta_\text{std}=0.13$\\
    \end{tabular}
    }}
  \end{overpic}
  \caption{
    In this Armadillo model, the original mesh has 83,699 vertices and 167,394 faces.
    \emph{Left}: If we do not use a simplification, then it takes 45.39s to compute the distortion points.
    \emph{Center}: The simplified mesh has 13,000 vertices and 25,996 faces. The distortion points are computed on this simplified mesh.
    \emph{Right}: The distortion points are mapped back onto the original mesh. The final distortion is similar, and this process takes only 20.37s.
  }
  \label{fig:simplification}
\end{figure}

\subsection{Implementation details}\label{sec:details}
\emph{Mesh Resolution.}
Since we compute ACAP parametrizations many times, this process is very time-consuming for large-scale models.
If a low-resolution mesh $\mM^s$ approximates the high-resolution model $\mM^l$ well and $\mM^s$ and $\mM^l$ are cut to disk topologies by similar cut paths, the distortion distributions of their planar parameterizations are similar as well.
As a result, the positions of the distortion points detected by $\mM^s$ are similar to the positions detected by $\mM^l$.
Fig.~\ref{fig:simplification} shows an example.
In our experiments, if $N_v > N_v^\text{thres}$, our detection algorithm contains three steps:
\begin{enumerate}
  \item Simplify $\mM$ using QEM~\cite{Garland:1997} to contain $N_v^\text{thres}$ vertices;
  \item Detect the distortion points of the simplified mesh;
  \item For each detected distortion point, find the nearest vertex on $\mM$, which becomes the resulting distortion point for $\mM$.
\end{enumerate}
We set $ N_v^\text{thres} = 13000$ in all examples.
If the threshold $N_v^\text{thres}$ is too small, some features may be missed.
Fig.~\ref{fig:parameters}-left shows an example of $N_v^\text{thres}=5000$.
Note that the top of the octopus is missing, leading to high distortion.

We consider the simplification as an optional operation that aims to accelerate the computation.
In our experiments, the simplification is not applied everywhere.
Specifically, if the mesh has a small number of vertices, there is no need to simplify the mesh.
Besides, if we do not simplify large-scale meshes, our method is still able to detect distortion points but takes more time.
Our experiments show that the simplification does not affect the robustness in practice.

\begin{figure}[!t]
  \centering
  \begin{overpic}[width=\linewidth]{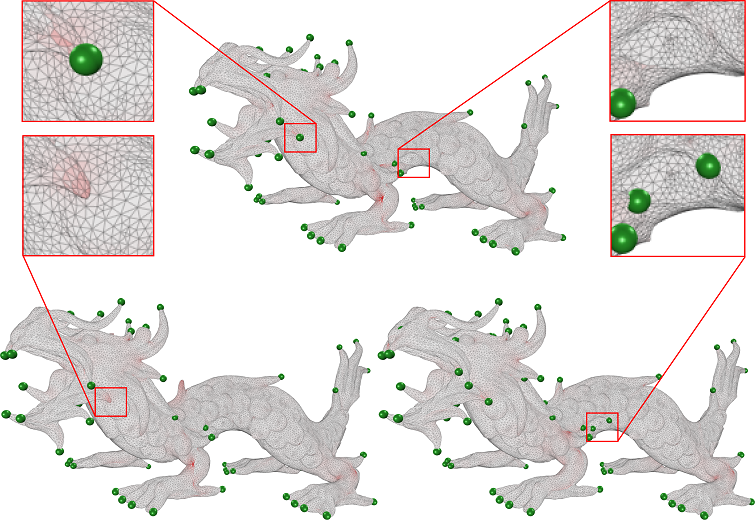}
    \put(21,26){\scriptsize \parbox{0.5\linewidth}{
    \begin{tabular}{l}
      $\boldsymbol{N=5}$\\
      \midrule
      $\delta_\text{avg}=1.16$\\
      $\delta_\text{max}=6.83$\\
      $\delta_\text{std}=0.17$\\
    \end{tabular}
    }}
    \put(46,62){\scriptsize \parbox{0.5\linewidth}{
    \begin{tabular}{l}
      $\boldsymbol{N=13}$\\
      \midrule
      $\delta_\text{avg}=1.16$\\
      $\delta_\text{max}=6.91$\\
      $\delta_\text{std}=0.17$\\
    \end{tabular}
    }}
    \put(71,26){\scriptsize \parbox{0.5\linewidth}{
    \begin{tabular}{l}
      $\boldsymbol{N=40}$\\
      \midrule
      $\delta_\text{avg}=1.18$\\
      $\delta_\text{max}=6.92$\\
      $\delta_\text{std}=0.19$\\
    \end{tabular}
    }}
  \end{overpic}
  \caption{
    Different values for $N$ generate different results.
    When $N$ is larger, there are fewer distortion points detected, and the distortion is larger.
  }
  \label{fig:diff_N}
\end{figure}

\emph{Choice of $N$.}
The parameter $N$ controls the balance between the number of distortion points and the isometric distortion.
A large $N$ value indicates that a distortion point has a large range of influence, so the number of distortion points is often small, resulting in high isometric distortion.
A small $N$ value usually leads to a large number of distortion points and small isometric distortion.
By default, we set $N = 0.1\%N_v$ if $N_v \leq N_v^\text{thres}$; otherwise, $N = 0.1\%N_v^\text{thres} = 13$.
Fig.~\ref{fig:diff_N} shows a comparison between different values for $N$.

\begin{figure}[!t]
  \centering
  \vspace{1mm}
  \begin{overpic}[width=\linewidth]{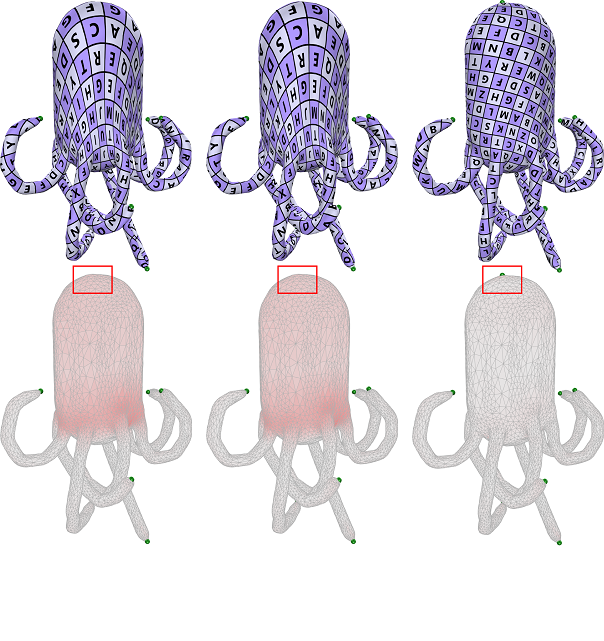}
    \put(0,6){\scriptsize \parbox{0.5\linewidth}{
    \begin{tabular}{p{2.3cm}}
      $\delta_\text{avg}=1.34$\\
      $\delta_\text{max}=4.94$\\
      $\delta_\text{std}=0.39$\\
      \midrule
      \textbf{Simplified to 5,000, 10 candidates}\\
    \end{tabular}
    }}
    \put(34,6){\scriptsize \parbox{0.5\linewidth}{
    \begin{tabular}{p{2.3cm}}
      $\delta_\text{avg}=1.35$\\
      $\delta_\text{max}=5.03$\\
      $\delta_\text{std}=0.39$\\
      \midrule
      \textbf{Simplified to 13,000, 5 candidates}\\
    \end{tabular}
    }}
    \put(66,6){\scriptsize \parbox{0.5\linewidth}{
    \begin{tabular}{p{2.3cm}}
      $\delta_\text{avg}=1.14$\\
      $\delta_\text{max}=5.10$\\
      $\delta_\text{std}=0.16$\\
      \midrule
      \textbf{Simplified to 13,000, 10 candidates}\\
    \end{tabular}
    }}
  \end{overpic}
  \caption{
    Results from different choices with other parameters.
    \emph{Left}: If the simplified mesh has too few vertices, then the distortion point at the top of the octopus is not detected (in the red rectangle).
    \emph{Middle}: If there are five candidate distortion detection processes, the distortion point at the top is not detected either.
    \emph{Right}: Our default choice, simplified to 13,000 vertices and ten candidates, finds the top distortion points successfully.
  }
  \label{fig:parameters}
\end{figure}

\emph{Parameter Selection.}
There are some parameters in the voting strategy.
We run ten candidate generation processes to generate candidates and select the candidate points that get at least three votes as distortion points.
The reasons for these settings are based on three cases of candidate points.
First, the candidate distortion points that are near the random cut have at most two votes (e.g. the points near a cut in Fig.~\ref{fig:pipeline}a2 top).
These points are caused by the cut and should not be involved in the final distortion points.
Second, when some candidate distortion points are located very close together (as shown in the red box in Fig.~\ref{fig:post}), they are not detected at the same time in one candidate generation process.
We observe that there are usually two or three candidate distortion points that are closely located in our experiments.
This case should be distinguished from the first case.
Third, the isolated candidate distortion points get almost all of the votes.
Therefore, in ten candidate generation processes, the candidate distortion points typically get one or two votes for the first case, four or five votes for the second case, and eight to ten votes for the third case.
On one hand, if there are fewer candidate generation processes, the points in the first and second situations cannot be distinguished.
On the other hand, it is feasible to conduct more candidate generation processes, but the entire algorithm requires more time to yield the final result.
We show a result with five candidates in the middle of Fig.~\ref{fig:parameters}.
In this example, the algorithm does not find the point at the top of the model and thus producing a high-distortion result.

\section{Experiments and Evaluations}\label{sec:results}
We apply our distortion point detection method to various models.
In Section~\ref{sec:eval}, we first evaluate the results and components of our algorithm and then demonstrate the practical robustness of our algorithm on different tessellations and complex models.
We select the Seamster~\cite{sheffer2002seamster}, Geometry Images method~\cite{Gu:2002}, and a sphere-based clustering method~\cite{ChaiCut2018} as the competitors (Section~\ref{sec:cmp}).
The experiments were performed on a desktop PC with an Intel Core i7-4790 processor and 8GB memory.

\begin{figure}[t]
  \centering
  \begin{overpic}[width=\linewidth]{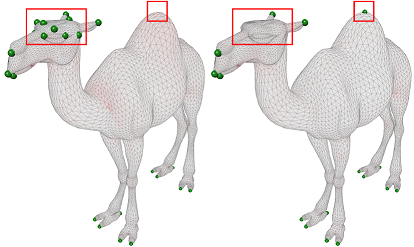}
    \put(0,13){\scriptsize \parbox{0.5\linewidth}{
    \begin{tabular}{p{2.1cm}}
      \textbf{Persistence-based method~\cite{chazal2013persistence}}\\
      \midrule
      $\delta_\text{avg}=1.17$\\
      $\delta_\text{max}=5.17$\\
      $\delta_\text{std}=0.17$\\
    \end{tabular}
    }}
    \put(50,11.4){\scriptsize \parbox{0.5\linewidth}{
    \begin{tabular}{l}
      \textbf{Ours}\\
      \midrule
      $\delta_\text{avg}=1.15$\\
      $\delta_\text{max}=4.60$\\
      $\delta_\text{std}=0.15$\\
    \end{tabular}
    }}
  \end{overpic}
  \caption{
    Comparison with the persistence-based clustering method~\cite{chazal2013persistence}.
    This method detects too many points on the camel's head and fails to find the distortion point on its hump, which results in higher distortion.
  }
  \label{fig:Persistence}
\end{figure}

\emph{Qualitative and Quantitative Evaluation Methods.}
Since the main purpose of finding distortion points is to decrease the distortion in parameterizations, a simple quantitative evaluation method is to conduct parameterizations and then compute the resulting distortion distribution.
Specifically, we adopt the isometric distortion metric defined in~\cite{Fu2015} to evaluate the quality of the parameterizations.
Then we report the distortion distribution by offering the maximum (worst case), average, and standard deviations for all triangles, denoted as $\delta_\text{max}$, $\delta_\text{avg}$, and $\delta_\text{std}$, respectively.
Meanwhile, we visualize the distortion distribution by shading the triangles according to the isometric distortion, with white being optimal (See Fig.~\ref{fig:pipeline}b top).
In addition, another popular visualization method we adopt is rendering mappings using a checkerboard texture with letters (See Fig.~\ref{fig:pipeline}b bottom).

\subsection{Evaluations}\label{sec:eval}

\begin{figure}[t]
  \centering
  \begin{overpic}[width=\linewidth]{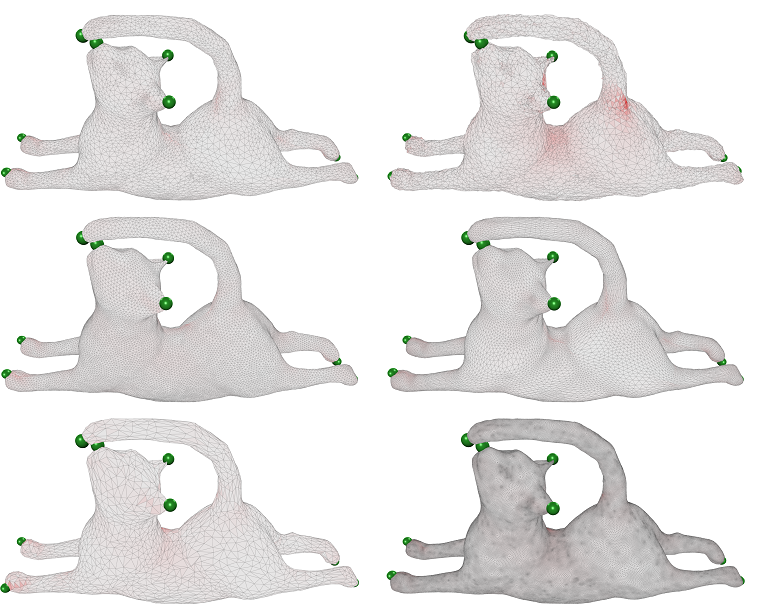}
    \put(30,70){\scriptsize \parbox{0.5\linewidth}{
    \begin{tabular}{r}
      \textbf{Original}\\
      \midrule
      $\delta_\text{avg}=1.13$\\
      $\delta_\text{max}=5.12$\\
      $\delta_\text{std}=0.14$\\
    \end{tabular}
    }}
    \put(30,44){\scriptsize \parbox{0.5\linewidth}{
    \begin{tabular}{r}
      \textbf{Isotropic}\\
      \midrule
      $\delta_\text{avg}=1.12$\\
      $\delta_\text{max}=4.26$\\
      $\delta_\text{std}=0.11$\\
    \end{tabular}
    }}
    \put(80,44){\scriptsize \parbox{0.5\linewidth}{
    \begin{tabular}{r}
      \textbf{Anisotropic}\\
      \midrule
      $\delta_\text{avg}=1.15$\\
      $\delta_\text{max}=6.26$\\
      $\delta_\text{std}=0.16$\\
    \end{tabular}
    }}
    \put(80,70){\scriptsize \parbox{0.5\linewidth}{
    \begin{tabular}{r}
      \textbf{Noisy}\\
      \midrule
      $\delta_\text{avg}=1.24$\\
      $\delta_\text{max}=6.26$\\
      $\delta_\text{std}=0.24$\\
    \end{tabular}
    }}
    \put(30,18){\scriptsize \parbox{0.5\linewidth}{
    \begin{tabular}{r}
      \textbf{Sparse}\\
      \midrule
      $\delta_\text{avg}=1.15$\\
      $\delta_\text{max}=3.63$\\
      $\delta_\text{std}=0.14$\\
    \end{tabular}
    }}
    \put(80,18){\scriptsize \parbox{0.5\linewidth}{
    \begin{tabular}{r}
      \textbf{Dense}\\
      \midrule
      $\delta_\text{avg}=1.15$\\
      $\delta_\text{max}=8.50$\\
      $\delta_\text{std}=0.15$\\
    \end{tabular}
    }}
  \end{overpic}
  \caption{
    Our method is robust to different mesh tessellations such as noise (right top), isotropy (left middle) or anisotropy (right middle), as well as low (left bottom) and high resolution (right bottom).
    The anisotropic mesh is generated by the LCT method~\cite{Fu2014}.
  }
  \label{fig:tess}
\end{figure}

\begin{figure*}[t]
  \centering
  \begin{overpic}[width=0.95\linewidth]{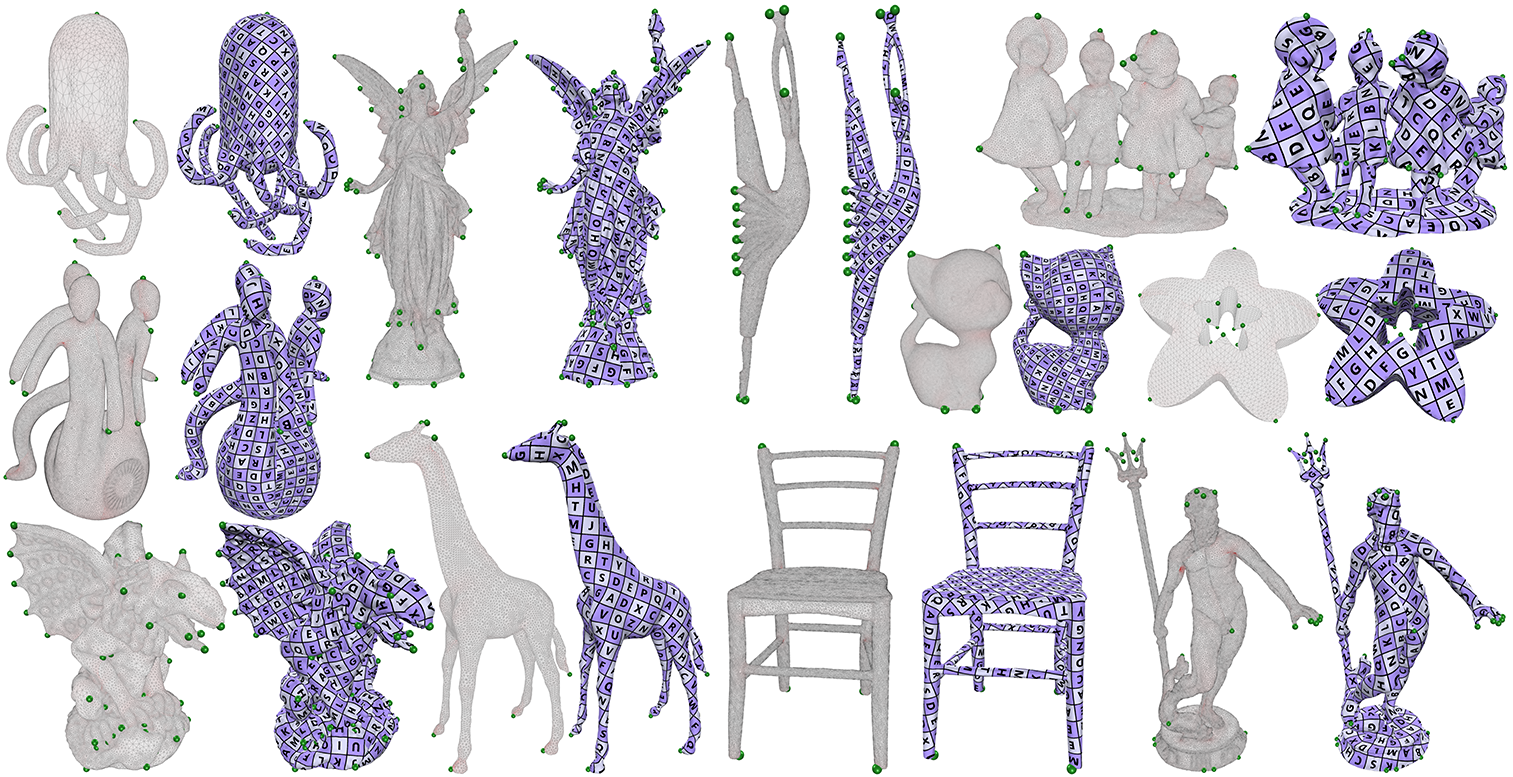}
  \end{overpic}
  \caption{
    Gallery displaying our distortion point detection results.
  }
  \label{fig:gallary}
\end{figure*}

\emph{Other Clustering Methods.}
Other existing clustering methods detect points according to the distortion of a parameterization.
We select one typical persistence-based method~\cite{chazal2013persistence} to compare with the effectiveness of our method.
As shown in Fig.~\ref{fig:Persistence}, the distortion points detected by the persistence-based clustering method are inappropriate, i.e., some points are superfluous, but some important regions are not detected.
In fact, the persistence-based method is regarded as a top-down approach, which finds the most significant features according to the topology persistence of the distortion.
However, our clustering method is a bottom-up approach, which filters out low distortion areas hierarchically.
As seen in the comparison, our bottom-up approach is more suitable for finding points that cause high distortion.

\emph{Tessellations.}
In Fig.~\ref{fig:tess}, distortion point detection is performed for six types of tessellations that represent the same shape.
The distortion points detected by our method are in very similar places, and the isometric distortions of the parameterizations are all at a low level.
Since the final distortion points are obtained by voting from several parameterizations, our method exhibits reliable results when using different tessellations as inputs.

\emph{Practical Robustness.}
Our method successfully detects the desired distortion points, resulting in parameterizations with low isometric distortion in all of the models.
We show eleven complex models in Fig.~\ref{fig:gallary}, including five genus-zero models and six high-genus models.
The C++ implementation for detecting distortion points for genus-zero surfaces is provided in the supplementary material.

\emph{Timings.}
The running time for distortion point detection on an ant model with 10k vertices is 8.64 seconds.
For a dragon model with 50k vertices, the mesh simplification and the distortion point detection take 3.44 seconds and 14.06 seconds, respectively.
Thanks to the use of mesh simplification, our method takes about 15 seconds on average.
The more structurally complex the input model is, the longer our method will take since more iterations are required to compute ACAP parameterizations.

\begin{figure}[!t]
  \centering
  \begin{overpic}[width=\linewidth]{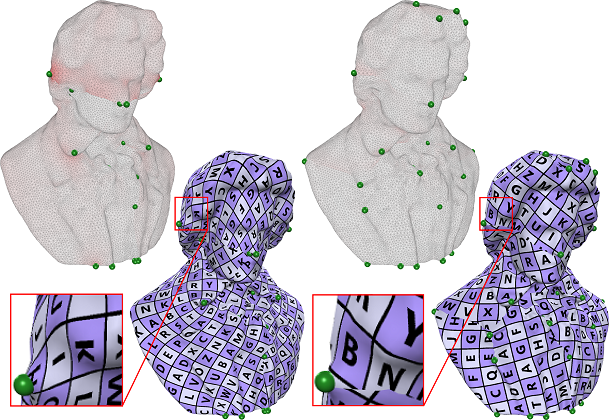}
    \put(30,55){\scriptsize \parbox{0.5\linewidth}{
    \begin{tabular}{l}
      \textbf{Seamster~\cite{sheffer2002seamster}}\\
      \midrule
      $\delta_\text{avg}=1.26$\\
      $\delta_\text{max}=9.56$\\
      $\delta_\text{std}=0.21$
    \end{tabular}
    }}
    \put(80,55){\scriptsize \parbox{0.5\linewidth}{
    \begin{tabular}{l}
      \textbf{Ours}\\
      \midrule
      $\delta_\text{avg}=1.10$\\
      $\delta_\text{max}=3.71$\\
      $\delta_\text{std}=0.10$
    \end{tabular}
    }}
  \end{overpic}
  \caption{
    Comparison to the Seamster method~\cite{sheffer2002seamster} using a Beethoven model.
    \emph{Left}: The Seamster method always finds high-curvature points, but in this example, some low curvature points also cause high distortion.
    \emph{Right}: Our method directly detects distortion points from planar parameterizations and therefore generates a result with lower distortion.
  }
  \label{fig:Seamster}
\end{figure}

\begin{figure}[!t]
  \centering
  \begin{overpic}[width=\linewidth]{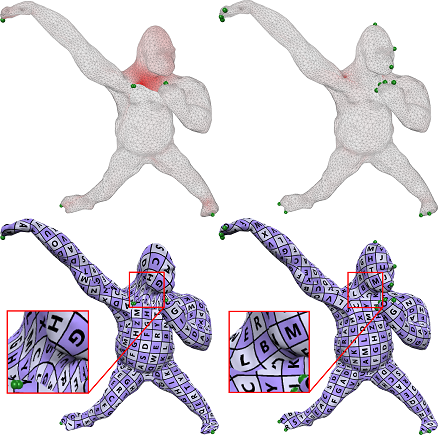}
    \put(4,70){\scriptsize \parbox{0.5\linewidth}{
    \begin{tabular}{p{1.4cm}}
      \textbf{Geometry Images~\cite{Gu:2002}}\\
      \midrule
      $\delta_\text{avg}=1.28$\\
      $\delta_\text{max}=6.90$\\
      $\delta_\text{std}=0.31$
    \end{tabular}
    }}
    \put(54,68){\scriptsize \parbox{0.5\linewidth}{
    \begin{tabular}{l}
      \textbf{Ours}\\
      \midrule
      $\delta_\text{avg}=1.14$\\
      $\delta_\text{max}=4.37$\\
      $\delta_\text{std}=0.13$
    \end{tabular}
    }}
  \end{overpic}
  \caption{
    Comparison with the Geometry Images~\cite{Gu:2002} method using a gorilla model.
    \emph{Left}: The result generated by the Geometry Images method has higher distortion since some important points are not detected.
    \emph{Right}: Our result has lower distortion.
  }
  \label{fig:geoImage}
\end{figure}

\begin{figure}[t]
  \centering
  \begin{overpic}[width=\linewidth]{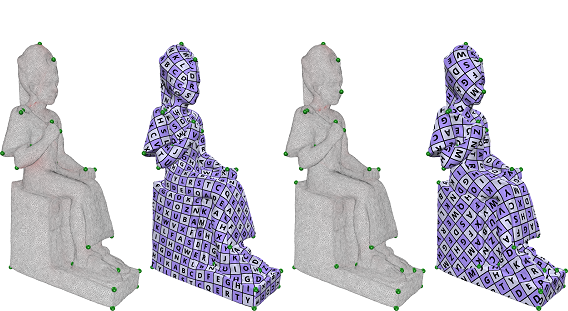}
    \put(9,45){\scriptsize \parbox{0.5\linewidth}{
    \begin{tabular}{l}
      \textbf{Sphere-based~\cite{ChaiCut2018}}\\
      \midrule
      $\delta_\text{avg}=1.13$\\
      $\delta_\text{max}=5.88$\\
      $\delta_\text{std}=0.15$
    \end{tabular}
    }}
    \put(59,45){\scriptsize \parbox{0.5\linewidth}{
    \begin{tabular}{l}
      \textbf{Ours}\\
      \midrule
      $\delta_\text{avg}=1.09$\\
      $\delta_\text{max}=5.50$\\
      $\delta_\text{std}=0.10$
    \end{tabular}
    }}
  \end{overpic}
  \caption{
    Comparison with the sphere-based method~\cite{ChaiCut2018} using a Ramses model. The input model contains 50,002 vertices and 100,000 triangles.
    The running times for the sphere-based method and our method are 42.75 seconds and 17.92 seconds, respectively.
    The simplification is enabled in our method.
  }
  \label{fig:sphere}
\end{figure}

\subsection{Comparisons}\label{sec:cmp}

\subsubsection{Comparison to Seamster~\cite{sheffer2002seamster}}
The Seamster method uses curvatures as guides for detecting distortion points, so the vertices with low curvatures are not recognized.
However, some low curvature vertices are important for reducing isometric distortion.
In Figs.~\ref{fig:curvature} and~\ref{fig:Seamster}, we compare our method to the Seamster method.
Due to some missing distortion points with low curvatures, the parameterizations generated by the Seamster method often have greater distortion than ours.

\subsubsection{Comparison to Geometry Images~\cite{Gu:2002}}
In Figs.~\ref{fig:distortion} and~\ref{fig:geoImage}, comparisons between our method and the Geometry Images method are conducted.
Due to early termination issues, some important points are missed, e.g., the points at the top of the human head (Fig.~\ref{fig:distortion}) and the gorilla head (Fig.~\ref{fig:geoImage}).
Thus, the parameterization distortion caused by the Geometry Images method is usually higher than ours.

\subsubsection{Comparison to the Sphere-Based Method~\cite{ChaiCut2018}}
In Fig.~\ref{fig:sphere}, we compare our method with the sphere-based method~\cite{ChaiCut2018}.
Note that since we use a different method to compute the final parameterizations, the distortion values are different from those in~\cite{ChaiCut2018}.
The AQP method~\cite{Kovalsky2016} and the SA method~\cite{Fu-2016-SA} are used in~\cite{ChaiCut2018}, while the progressive parameterization method~\cite{LiuPP-2018} is used in this paper.
Thus, we recompute a parameterization using the same method and measure the distortion for fair comparison.
The detected distortion points are similar, so the parameterization distortion is also similar.
However, as the bijective spherical parameterization takes 42.09 seconds, the sphere-based method takes more time than our method.

\subsubsection{Extensive Comparisons}
\emph{Data Set Construction.}
To verify the robustness of our method and compare it with other methods, we construct a data set containing 20,000 random genus-zero meshes.
The meshes in the data set are constructed from a primitive shape, such as a sphere, a cube, a pyramid, a dodecahedron, etc.
Then, we conduct random shape manipulations, including rotation, scaling, random affine transformation, stretching, bending, twisting, random vertex perturbation, fractal displacement~\cite{ebert2003texturing}, CSG manipulations (merge, intersect, and subtract with other models), remeshing, smoothing, and simplification~\cite{botsch2010polygon}.
After each manipulation, if the mesh is not a genus zero manifold, we discard this manipulation and roll back to the previous shape.
Each mesh is finally simplified or subdivided to 13,000 vertices. We show five examples in Fig.~\ref{fig:dataset}.

\begin{figure}[t]
  \centering
  \begin{overpic}[width=\linewidth]{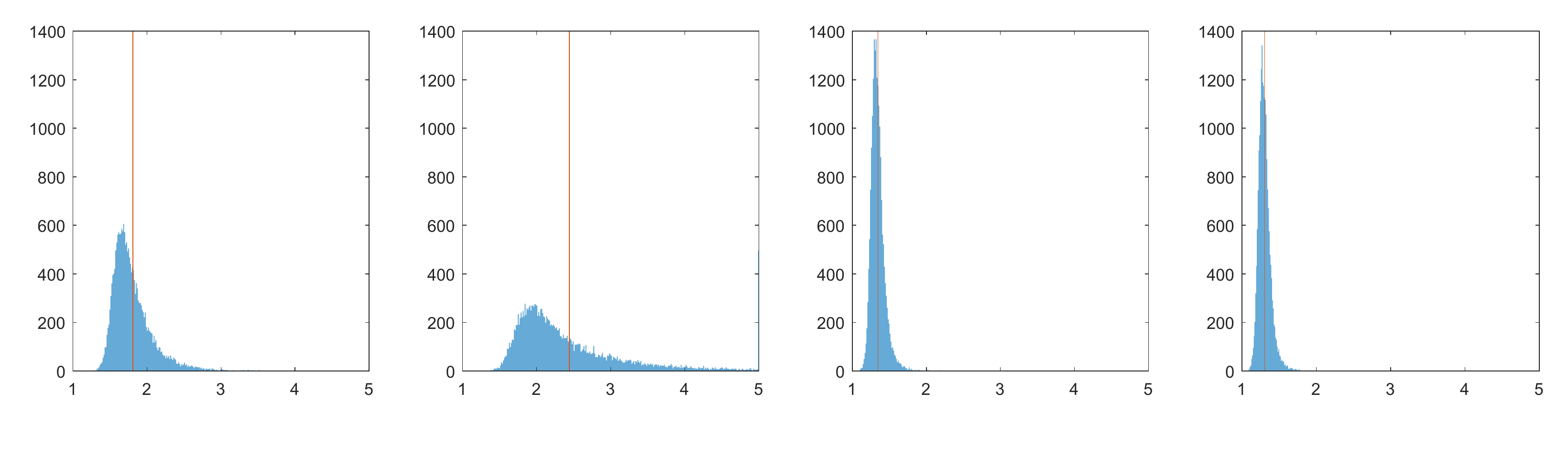}
    \put(9,24){\scriptsize 1.81}
    \put(37,24){\scriptsize 2.44}
    \put(57,24){\scriptsize 1.35}
    \put(82,24){\scriptsize 1.31}
    \put(0,0){\scriptsize\textbf{Geometry Images~\cite{Gu:2002}}}
    \put(31,0){\scriptsize\textbf{Seamster~\cite{sheffer2002seamster}}}
    \put(52,0){\scriptsize\textbf{Sphere-based~\cite{ChaiCut2018}}}
    \put(85,0){\scriptsize\textbf{Ours}}
    \put(94,2){\scriptsize$\delta_\text{avg}$}
    \put(0,28){\scriptsize \#model}
  \end{overpic}
  \caption{
    The histogram showing the average distortion in the results from four algorithms running on 20,000 models.
    The red lines with numbers show the average values.
  }
  \label{fig:stat}
\end{figure}

\begin{figure*}[t]
  \centering
  \includegraphics[width=0.95\linewidth]{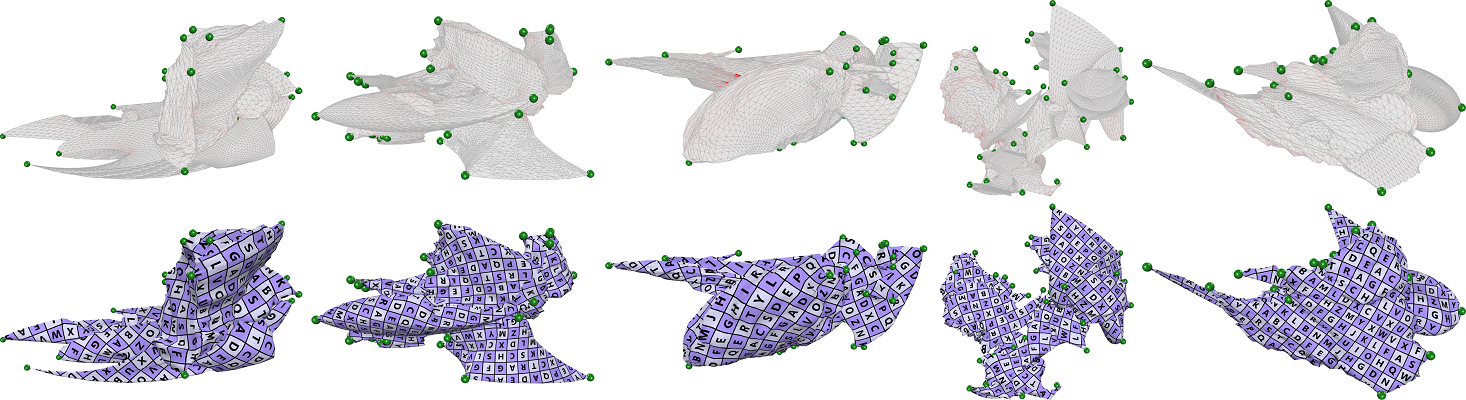}
  \caption{
     Detected distortion points in five models from the constructed data set.
  }
  \label{fig:dataset}
\end{figure*}

\emph{Results.}
We have run the Geometry Images method~\cite{Gu:2002}, the Seamster method~\cite{sheffer2002seamster}, the sphere-based method~\cite{ChaiCut2018}, and our method on the constructed data set.
The Seamster method, the Geometry Images method, and our method succeed in all models, while the sphere-based method fails in 508 models.
The sphere-based method depends on a spherical parameterization.
If the input model has a long and thin neck, the sphere-based method~\cite{ChaiCut2018} usually fails due to its spherical parameterization method~\cite{Hu-2017-AHSP}.
Specifically, such a long and thin neck provides no enough space to insert vertices, so the spherical parameterization gets stuck in the vertex insertion phase.
Since the possible vertex positions are enumerated discretely by sampling points in a triangle, if flips occur for all vertex positions, the algorithm terminates and does not generate a flip-free spherical parameterization.
Fig.~\ref{fig:dataset} shows its five failed examples.

The histograms in Fig.~\ref{fig:stat} show the distributions of $\delta_\text{avg}$.
The average $\delta_\text{avg}$ over all successful examples of~\cite{ChaiCut2018} is higher than our method.
Since some important points are not detected by the Seamster method~\cite{sheffer2002seamster} and Geometry Images method~\cite{Gu:2002}, their average $\delta_\text{avg}$ are much higher than ours.

\section{Applications}\label{sec:app}
We employ our detected distortion points in three applications.

\begin{figure}[t]
  \centering
  \vspace{1mm}
  \begin{overpic}[width=\linewidth]{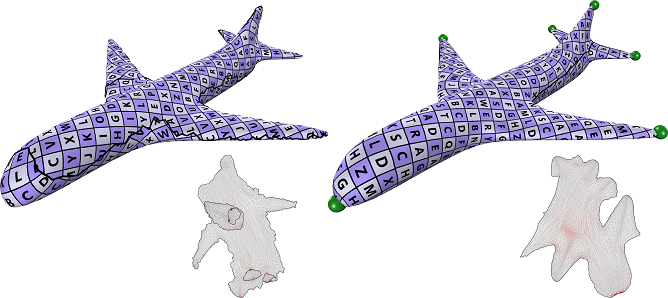}
    \put(5,9){\scriptsize \parbox{0.5\linewidth}{
    \begin{tabular}{r}
      \textbf{Autocuts}\\
      \midrule
      $\delta_\text{avg}=1.15$\\
      $\delta_\text{max}=3.95$\\
      $\delta_\text{std}=0.23$
    \end{tabular}
    }}
    \put(55,9){\scriptsize \parbox{0.5\linewidth}{
    \begin{tabular}{r}
      \textbf{Ours}\\
      \midrule
      $\delta_\text{avg}=1.17$\\
      $\delta_\text{max}=4.67$\\
      $\delta_\text{std}=0.20$
    \end{tabular}
    }}
  \end{overpic}
  \caption{
    Comparison with Autocuts~\cite{poranne2017autocuts}.
    \emph{Left}: The Autocuts method generates cuts and parameterizations simultaneously.
    \emph{Right}: We generate a cut using a spanning tree connecting the distortion points.
    The ratio between the cut lengths and total edge lengths in the Autocuts method and ours are $5.27\%$ and $2.07\%$.
  }
  \label{fig:autocut}
\end{figure}

\begin{figure}[t]
  \centering
  \begin{overpic}[width=\linewidth]{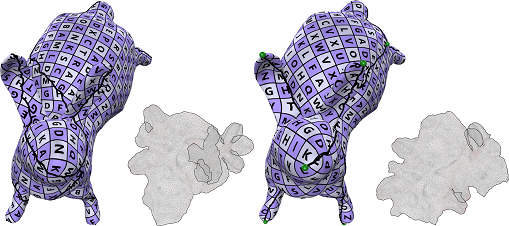}
    \put(28,33){\scriptsize \parbox{0.5\linewidth}{
    \begin{tabular}{r}
      \textbf{OptCuts}\\
      \midrule
      $\delta_\text{avg}=1.15$\\
      $\delta_\text{max}=6.14$\\
      $\delta_\text{std}=0.14$
    \end{tabular}
    }}
    \put(78,33){\scriptsize \parbox{0.5\linewidth}{
    \begin{tabular}{r}
      \textbf{Ours}\\
      \midrule
      $\delta_\text{avg}=1.14$\\
      $\delta_\text{max}=5.08$\\
      $\delta_\text{std}=0.15$
    \end{tabular}
    }}
  \end{overpic}
  \caption{
    Comparison with OptCuts~\cite{Li:2018:OptCuts}.
    \emph{Left}: The result is generated by the Optcuts method.
    \emph{Right}: We generate a cut using a spanning tree connecting the distortion points.
    The ratio between the cut lengths and total edge lengths in the OptCuts method and ours are $1.27\%$ and $1.00\%$.
  }
  \label{fig:optcuts}
\end{figure}

\subsection{Parameterizations}
Computing parameterizations is a straightforward application of our method.
Given a closed mesh, we generate a cut by simply constructing a minimal spanning tree along the mesh edges that passes through all the distortion points detected by our method.
After cutting the mesh into a disk topology, any of the recent parameterization methods are able to generate a low-distortion UV map, which is used for texture mapping.
In our experiments, we use the progressive parameterizations method~\cite{LiuPP-2018} to generate UV maps and color the triangles according to the isometric distortion with red indicating high distortion.
In addition, we also use a checkerboard texture with letters to show the levels of distortion.

In Figs.~\ref{fig:autocut} and~\ref{fig:optcuts}, we give a brief comparison with the recent Autocuts method~\cite{poranne2017autocuts} and OptCuts method~\cite{Li:2018:OptCuts}.
We only adjust the parameters $\lambda$ and $\delta$ in the Autocuts method to obtain results with similar distortion levels to ours.
For the OptCuts method, we turn off the bijective option and also adjust the parameters to obtain similar distortion results.
Then, we compare the cut lengths with these two methods.
The cut lengths of our result are shorter than theirs.
Moreover, our method is fully automatic and does not require any user intervention.

\subsection{Semi-Automatic Landmark Correspondence}
\begin{figure}[t]
  \centering
  \begin{overpic}[width=\linewidth]{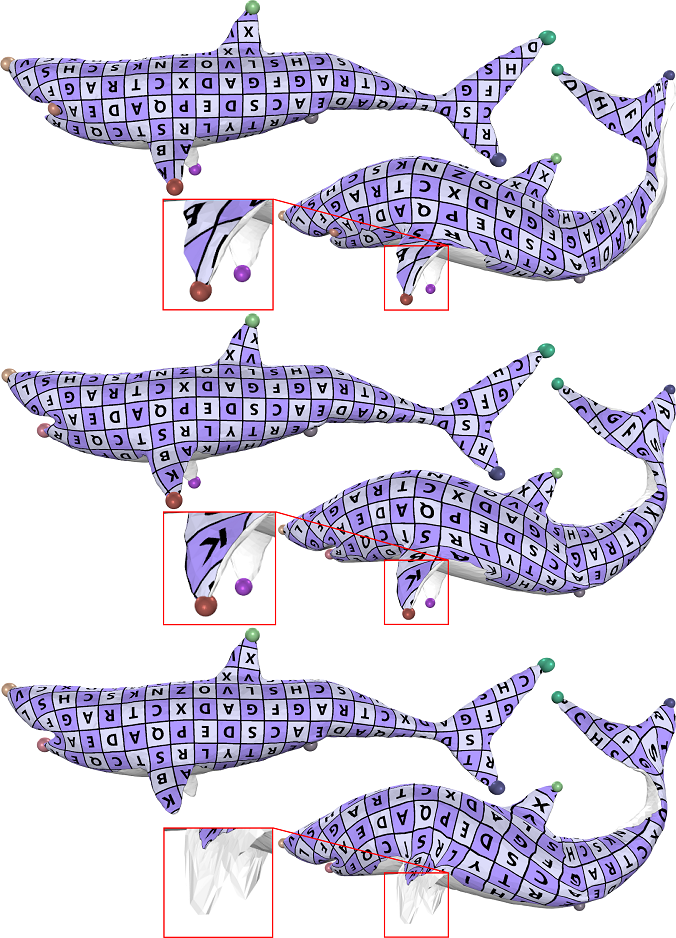}
    \put(0,73){\scriptsize \parbox{0.5\linewidth}{
    \begin{tabular}{p{1.4cm}}
      \textbf{Manually Picking}\\
      \midrule
      $\delta_\text{avg}=1.65$\\
      $\delta_\text{max}=26.2$\\
      $\delta_\text{std}=0.47$
    \end{tabular}
    }}
    \put(0,40){\scriptsize \parbox{0.5\linewidth}{
    \begin{tabular}{p{1.4cm}}
      \textbf{With Distortion Points}\\
      \midrule
      $\delta_\text{avg}=1.62$\\
      $\delta_\text{max}=31.1$\\
      $\delta_\text{std}=0.46$
    \end{tabular}
    }}
    \put(0,7){\scriptsize \parbox{0.5\linewidth}{
    \begin{tabular}{p{1.4cm}}
      \textbf{Missing Two Points}\\
      \midrule
      $\delta_\text{avg}=1.93$\\
      $\delta_\text{max}=21.4$\\
      $\delta_\text{std}=0.74$
    \end{tabular}
    }}
  \end{overpic}
  \caption{
    The inter-surface mapping between two fish models using landmark correspondences as inputs. Here, the distortions shown in the figure are the isometric distortions of the inter-surface mappings.
    \emph{Top}: The locations and corresponding landmarks are selected by hand.
    \emph{Center}: The landmarks are the distortion points generated by our method and their correspondence is specified manually.
    \emph{Bottom}: If there are some missed landmarks, the distortion is large.
    The manually selected landmarks are provided by~\cite{YangCM2018}.
  }
  \label{fig:correspondence}
\end{figure}

Given two meshes, we first generate the distortion points for each mesh and use these points as corresponding landmarks, where the correspondence is manually specified.
Then, we find cuts that pass through these landmarks and use the lifted bijection method~\cite{Aigerman2014} to compute a bijective surface mapping.
We give an example of this application in Fig.~\ref{fig:teaser} center and a comparison in Fig.~\ref{fig:correspondence}.
From the comparison, we find that the locations of distortion points (center) and manually selected landmarks (top) are similar, and the distortions of the resulting inter-surface mappings are also very similar.
However, if some of the distortion points are missing (bottom), the distortion becomes dramatically larger.
Therefore, our method has the ability to automatically detect the most important features in models and reduce user interactions when generating inter-surface mappings.

\begin{figure}[t]
  \centering
  \begin{overpic}[width=\linewidth]{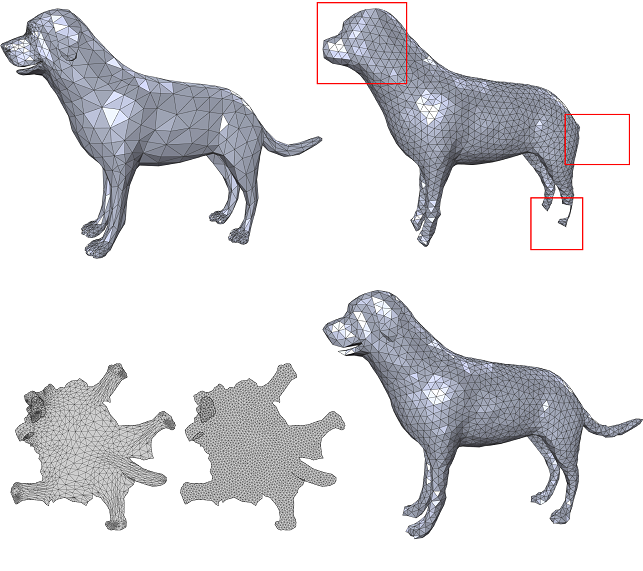}
    \put(21,45){\scriptsize\textbf{Input}}
    \put(58,45){\scriptsize\textbf{Incremental remeshing\cite{botsch2004remeshing}}}
    \put(2,1){\scriptsize\textbf{Parameterization}}
    \put(31,1){\scriptsize\textbf{2D remeshing}}
    \put(68,1){\scriptsize\textbf{Our result}}
  \end{overpic}
  \caption{
    Isotropic remeshing of a dog model.
    \emph{Left top}: The original mesh has some thin parts (e.g. the legs and the tail) and some narrow ditches (e.g. the mouth and the ears).
    \emph{Right top}: The incremental remeshing method~\cite{botsch2004remeshing} smooths out the thin features (shown in the red box).
    \emph{Left bottom}: Our method parameterizes the model to the plane isometrically and does isotropic remeshing on the planar mesh.
    \emph{Right bottom}: After mapping the 2D isotropic mesh back to the original model, the thin features are correctly preserved.
  }
  \label{fig:remesh}
\end{figure}
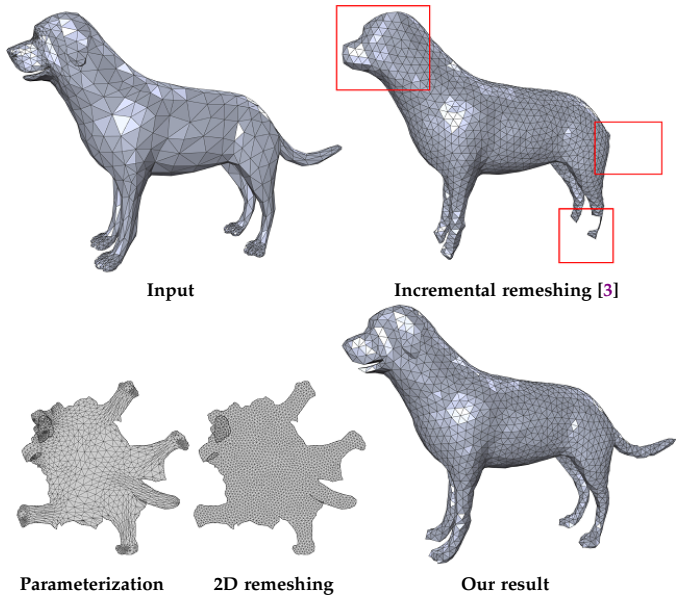

\subsection{Isotropic Remeshing}
The incremental remeshing method~\cite{botsch2004remeshing} is an efficient and effective algorithm for generating isotropic triangle meshes.
Given a target edge length, this method repeatedly carries out four steps:
(1) splitting long edges,
(2) collapsing short edges,
(3) equalizing valences using edge flipping,
and (4) relocating vertices.
However, certain features, such as thin parts and narrow ditches, are often collapsed (see Fig.~\ref{fig:remesh} right top).
When a cut passes through the distortion points, we usually get a parameterization with very low isometric distortion.
Since the mapping from the original surface to the parameterization domain is able to preserve isotropy, we do an isotropic remeshing on the parameterization mesh.
First, we split and collapse edges on the cut before the parameterization, and fix the boundary when remeshing the 2D parameterization mesh.
Then, the four aforementioned steps in the incremental remeshing method are carried out to create a 2D isotropic mesh, which is mapped back onto the original mesh.
Fig.~\ref{fig:remesh} shows a comparison of three dog models.
The thin features collapse in the incremental remeshing method~\cite{botsch2004remeshing}, while our method preserves these features successfully.
The reason is that the edge lengths in~\cite{botsch2004remeshing} are computed using the 3D Euclidean distance, which yields short lengths at the thin parts; thus, the collapsing operations destroy these thin features.
However, the edge lengths in our method are computed using 2D Euclidean distances that approximate the geodesic distances of an input mesh, which is able to preserve thin features. This characteristic is very useful in the error-bounded remeshing~\cite{Cheng2019}.

\section{Conclusion}\label{sec:conclusion}
In this paper, we present a novel approach for detecting distortion points on 3D triangle meshes.
This method can be further applied to planar parameterization, semi-automatic landmark correspondence, and isotropic remeshing.

Our method has some limitations.
First, there is no theoretical analysis regarding the relationships between the detected distortion points and the resulting parameterization distortion.
Our method is empirical, based on the observation that the isometric distortion will be low if the cuts pass through the distortion points.
Although cutting through the distortion points tends to generate low-distortion parameterizations, this condition is not always guaranteed.
For example, as shown in Fig.~\ref{fig:tess}, the noisy model has higher distortion than the others, even if the cut passes through the distortion points.
The second limitation is that for a sphere-shaped mesh, our method cannot find any distortion points.
The reason is that the vertices are almost all symmetrically equivalent, and the random cuts generate random candidate distortion points, each of which has only one vote.

\section*{Acknowledgments}
The authors would like to thank the anonymous reviewers for their constructive suggestions and comments.
This work is supported by the National Natural Science Foundation of China (61802359, 61672482, 11626253), the Anhui Provincial Natural Science Foundation (1808085QF208), and the Fundamental Research Funds for the Central Universities (WK0010460006, WK0010450004).

\ifCLASSOPTIONcaptionsoff
  \newpage
\fi

\bibliographystyle{IEEEtran}

%

\vskip -2\baselineskip plus -1fil
\begin{IEEEbiography}[{\includegraphics[width=1in,height=1.25in,clip,keepaspectratio]{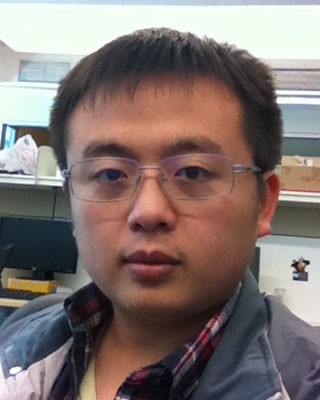}}]{Shuangming Chai}
received a BSc degree in 2014 from the University of Science and Technology of China. He is currently a PhD candidate at the School of Mathematical Sciences, University of Science and Technology of China. His research interests include geometric processing and 3D modeling.
\end{IEEEbiography}
\vskip -2\baselineskip plus -1fil
\begin{IEEEbiography}[{\includegraphics[width=1in,height=1.25in,clip,keepaspectratio]{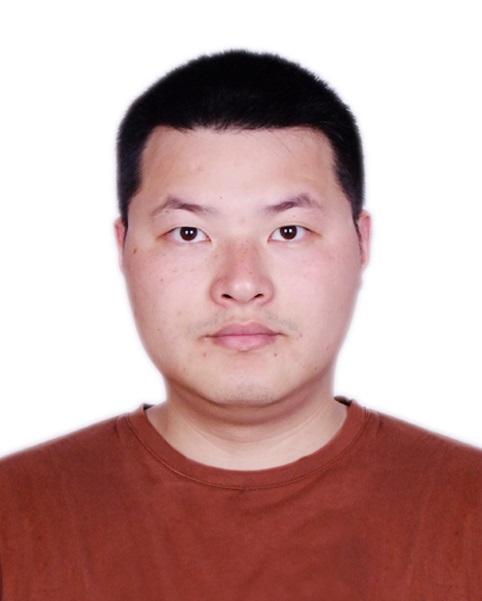}}]{Xiao-Ming Fu}
received a BSc degree in 2011 and a PhD degree in 2016 from University of Science and Technology of China. Currently, he is an assistant researcher at the School of Mathematical Sciences, University of Science and Technology of China. His research interests include geometric processing and computer-aided geometric design. His research work can be found at his research website: \url{http://staff.ustc.edu.cn/~fuxm}.
\end{IEEEbiography}
\vskip -2\baselineskip plus -1fil
\begin{IEEEbiography}[{\includegraphics[width=1in,height=1.25in,clip,keepaspectratio]{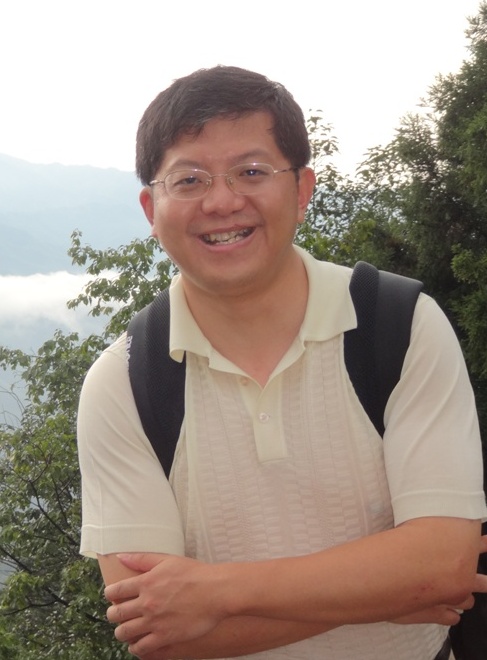}}]{Ligang Liu}
is a professor at the School of Mathematical Sciences, University of Science and Technology of China. He received his B.Sc. (1996) and his Ph.D. (2001) from Zhejiang University, China. Between 2001 and 2004, he worked at Microsoft Research Asia. Then, he worked at Zhejiang University during 2004 and 2012. He paid an academic visit to Harvard University during 2009 and 2011. His research interests include digital geometric processing, computer graphics, and image processing. He serves as the associate editors for the journals of IEEE Transactions on Visualization and Computer Graphics, IEEE Computer Graphics and Applications, Computer Graphics Forum, Computer Aided Geometric Design, and The Visual Computer. He served as the conference co-chair of GMP 2017 and the program co-chairs of GMP 2018, CAD/Graphics 2017, CVM 2016, SGP 2015, and SPM 2014. His research work can be found at his research website: \url{http://staff.ustc.edu.cn/~lgliu}.
\end{IEEEbiography}

\end{document}